\documentclass[12pt,aps]{article}
\usepackage{epsfig}
\usepackage{amsfonts}
\usepackage{mathrsfs}
\usepackage{amsmath}
\usepackage{amssymb}
\usepackage{cite}
\usepackage{xcolor}
\usepackage{epic}
\usepackage{tikz}
\usepackage[title]{appendix}
\usepackage{cite}

\begin{document}
\sffamily
\title{
Wave function of ${}^9\mathrm{Be}$ in the three-body $(\alpha\alpha n)$-model
}
\author{
S. A. Rakityansky$^{\,a,b}$\\[3mm] 
\parbox{11cm}{%
${}^a${\small Joint Institute for Nuclear Research, Dubna, Russia}\\
${}^b${\small Department of Physics, University of Pretoria, Pretoria,
South Africa}}
}
\maketitle
\begin{abstract}
\noindent
A simple analytic expression of the three-body wave function describing the 
system $(\alpha\alpha n)$ in the ground state $\frac{3}{2}^-$ of 
${}^9\mathrm{Be}$ is obtained. In doing this, it is assumed that the $\alpha$ 
particles interact with each other via the $S$-wave Ali-Bodmer potential 
including the Coulomb term, and the neutron-$\alpha$ forces act only in the 
$P$-wave state. This wave function is constructed by trial and error method via 
solving in this way a kind of inverse problem when the two-body $\alpha\alpha$ 
potential is recovered from a postulated three-body wave function. As a result, 
the wave function is an exact solution of the corresponding three-body 
Schr\"odinger equation for experimentally known binding energy and for the 
$\alpha\alpha$ potential whose difference from the Ali-Bodmer one is minimized 
by varying the adjustable parameters which the postulated wave function depends 
on.
\end{abstract}
\section{Introduction}
\label{sec.Introduction}
Beryllium isotope ${}^9\mathrm{Be}$ is of special interest from several points 
of view. First of all, it is clusterized with very high probability in two 
$\alpha$ particles and a neutron. As a result the low-lying spectrum of this 
nucleus is well-reproduced within the three-body ($\alpha\alpha n$)-model  
\cite{Tang, Efros, Filikhin, Casal, Filandri}. 

Secondly, neither $\alpha\alpha$ 
nor $\alpha n$ pairs can form bound states, which means that this isotope is an 
example of a Borromean system with a rather small binding energy of 
$1.5736\,\mathrm{MeV}$ \cite{Tilley}. 

Thirdly, due to such a weak three-body 
binding the valence neutron is loosely bound as well. This means that the 
neutron should move rather far away from the centre of mass, which makes the 
nucleus ${}^9\mathrm{Be}$ a transient one between ordinary compact nuclei and 
the halo-nuclei. 

And finally, the last but not least is the fact that 
${}^9\mathrm{Be}$ plays an important role in astrophysical processes, namely, in 
synthesis of heavy elements in the universe. Since there are no stable nuclei 
with $A=5$ and $A=8$, the two-body $pp$-chain reactions practically stop at the 
formation of ${}^4\mathrm{He}$ \cite{cauldrons, Aprahamian}. The synthesis of 
more heavy elements requires a bridge over these so-called $A=5,8$ mass gaps. 
These gaps can be crossed over via the three-body fusion reactions. Among them 
the most well known is the triple-alpha fusion,  
$\alpha\alpha\alpha\rightarrow {}^{12}\mathrm{C}+\gamma$, which is usually 
associated with the Hoyle resonance (see, for example, Ref. \cite{Khan}).  In 
a neutron-rich environment one of the alternative bridges is provided by the 
radiative process $n(\alpha\alpha,\gamma){}^9\mathrm{Be}$ which involves the 
Beryllium isotope that is considered in the present paper.

Of course, in addition to what was already said, the nucleus ${}^9\mathrm{Be}$ 
appears either in the initial or in the final states of various nuclear 
reactions, such as ${}^9\mathrm{Be}(n,\gamma){}^{10}\mathrm{Be}$ \cite{Mohr}, 
${}^9\mathrm{Be}({}^{18}\mathrm{O},{}^{17}\mathrm{O}){}^{10}\mathrm{Be}$ 
\cite{Carbone}, ${}^8\mathrm{Li}(p,\gamma){}^{9}\mathrm{Be}$ \cite{SuJun, Dub}, 
etc.

When theoretically describing the processes mentioned above, one usually 
needs the wave functions of the quantum systems involved, and in particular the 
wave function of  ${}^9\mathrm{Be}$. There are many different approaches to 
obtaining such a function. 

Strictly speaking, this nucleus should be considered 
as a nine-body system. Such an (ab initio) approach can be based either on the 
shell model 
or on the  expansion over the hyperspherical 
harmonics. 
However, as was mentioned before, ${}^9\mathrm{Be}$ is 
clusterized in two $\alpha$ particles and a neutron. Thanks to the tight binding 
of ${}^4\mathrm{He}$ ($\sim28\,\mathrm{MeV}$), the $\alpha$ particles move 
inside ${}^9\mathrm{Be}$ like solid bodies without internal excitations,  and 
therefore the nine-body problem can be reduced to an effective three-body one. 
To the best of the author's knowledge all the calculations that were done for  
${}^9\mathrm{Be}$ during the last few decades, exploited the $(\alpha\alpha 
n)$ cluster-representation. 
The three-body problem can also be solved in different ways: either using exact 
Faddeev equations (see, for example, Refs. \cite{Efros, Filikhin}), or with the 
help of various approximate methods \cite{Casal, Itagaki, Descouvemont1, 
Descouvemont}.

The common feature of all the publications, where these three-body 
approaches are realised, is that the wave function of ${}^9\mathrm{Be}$ is not 
given (and in most cases cannot be given) in such a form that could be used by 
the other people in their own calculations. In other words, if somebody wants 
to use the same wave function, he or she has to repeat all the complicated 
calculations described in these publications. However, for the majority of the 
researchers this is a difficult obstacle. Indeed, the procedure of solving, for 
example, the Faddeev equations involves many mathematical and numerical tricks 
that are only known to those who specialize in the field of the few-body 
problem. The shell model and the hyperspherical expansion are not much 
easier. It is therefore desirable to have some parametrized analytic 
expressions of the wave functions of various nuclei that are easy to use. This 
is what the present paper is devoted to. Here such a parametrization is 
obtained for the wave function of ${}^9\mathrm{Be}$ in the three-body cluster 
model $(\alpha\alpha n)$.

The way of obtaining the parametrized three-body wave function is based on the 
work by Belyaev et al. \cite{Belyaev}. That parer deals with a three-body 
problem in which two of the three pairwise potentials as well as the three-body 
wave function are given while one of the two-body potentials is unknown. The 
authors of Ref. \cite{Belyaev} developed a method for recovering this unknown 
potential.

In the present work, this method is kind of ``reversed'', i.e. the wave 
function is postulated in a parametrized form and its parameters are optimized 
by minimizing the difference between the recovered potential and the 
corresponding exact one. For the system under consideration, $(\alpha\alpha 
n)$, it is chosen the $\alpha\alpha$ potential for the role of recovered one. 
The analytic form of the wave function and its parameters are chosen to fit 
(as close as possible) the Ali-Bodmer potential \cite{AliBodmer}.

\section{Formalism}
\label{sec.formalism}
\begin{figure}[ht!]
\centerline{\epsfig{file=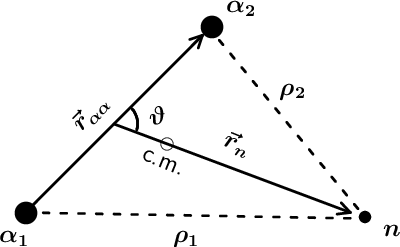}}
\caption{\sf
Jacobi coordinates, $\vec{r}_{\alpha\alpha}$ and $\vec{r}_n$, that specify a 
space configuration of the system $(\alpha\alpha n)$. The symbols $\rho_1$ and 
$\rho_2$ denote distances between the neutron and the two alpha particles.
}
\label{fig.Jacobi}
\end{figure}

Consider a bound state of two $\alpha$ particles and a neutron with a negative 
three-body energy $E_b$. It is known \cite{Tilley} that this system has only 
one bound state, which is the nucleus ${}^9\mathrm{Be}$. 
Using the Jacobi coordinates shown in Fig. \ref{fig.Jacobi}, the  
three-body Schr\"odinger equation for the wave function 
$\Psi(\vec{r}_{\alpha\alpha},\vec{r}_n)$ describing ${}^9\mathrm{Be}$, can be 
written as follows:
\begin{eqnarray}
\label{Schreq}
   && V_{\alpha\alpha}({r}_{\alpha\alpha})\Psi(\vec{r}_{\alpha\alpha},\vec{r}_n)
   =\\[3mm]
\nonumber
   &=&
   \left[
   E_b-V_{\alpha n}(\rho_1)-V_{\alpha n}(\rho_2)
   +\frac{\hbar^2}{2\mu_{\alpha\alpha}}
   \Delta_{\vec{r}_{\alpha\alpha}}+\frac{\hbar^2}{2\mu}
   \Delta_{\vec{r}_{n}}
   \right]
   \Psi(\vec{r}_{\alpha\alpha},\vec{r}_n)\ ,
\end{eqnarray}
where $V_{\alpha\alpha}$ and $V_{\alpha n}$ are the two-body potentials for the 
$\alpha\alpha$ and $\alpha n$ pairs, $\mu_{\alpha\alpha}=m_\alpha/2$ and
$\mu=2m_\alpha m_n/(2m_\alpha+m_n)$ are the reduced masses corresponding to the 
respective Jacobi coordinates, the radial variables $\rho_1$ and $\rho_2$ given 
by
\begin{equation}
\label{r12}
  \rho_{1,2}=\sqrt{\displaystyle\frac{1}{4}r_{\alpha\alpha}^2+r_n^2\mp
      r_{\alpha\alpha}r_n\cos\vartheta}\ ,
\end{equation}
are the distances between the neutron the the $\alpha$ particles, and the 
Laplacians involve the derivatives with respect to the corresponding Jacobi 
vectors. 

Multiplying Eq. (\ref{Schreq}) by 
$\Psi^\dagger(\vec{r}_{\alpha\alpha},\vec{r}_n)$ from the left and integrating 
over vector $\vec{r}_n$ as well as over the spherical angles, 
$\Omega_{\alpha\alpha}$,  of vector $\vec{r}_{\alpha\alpha}$, one can find the 
potential $V_{\alpha\alpha}$, if the energy $E_b$, the potential $V_{\alpha 
n}$, and the function $\Psi(\vec{r}_{\alpha\alpha},\vec{r}_n)$ are given,
\begin{eqnarray}
\label{Vgeneral}
   && V_{\alpha\alpha}({r}_{\alpha\alpha})
   =E_b-\\[3mm]
\nonumber
   &-&
   \frac{1}{D({r}_{\alpha\alpha})}
   \int d\Omega_{\alpha\alpha} d^3r_n
   \Psi^\dagger(\vec{r}_{\alpha\alpha},\vec{r}_n)\left[
   \!\!\phantom{\frac{|}{|}}
   V_{\alpha n}(\rho_1)+V_{\alpha n}(\rho_2)-\right.\\[3mm]
\nonumber
   &-&
   \left.\frac{\hbar^2}{2\mu_{\alpha\alpha}}
   \Delta_{\vec{r}_{\alpha\alpha}}-\frac{\hbar^2}{2\mu}
   \Delta_{\vec{r}_{n}}
   \right]
   \Psi(\vec{r}_{\alpha\alpha},\vec{r}_n)\ ,
\end{eqnarray}
where
\begin{equation}
\label{D}
   D({r}_{\alpha\alpha})=
   \int d\Omega_{\alpha\alpha} d^3r_n
   \Psi^\dagger(\vec{r}_{\alpha\alpha},\vec{r}_n)
   \Psi(\vec{r}_{\alpha\alpha},\vec{r}_n)
\end{equation}
is the relative $\alpha\alpha$ radial probability density.

It should be noted that the wave function 
$\Psi(\vec{r}_{\alpha\alpha},\vec{r}_n)$ is (by construction) the exact solution 
of the Schr\"odinger equation with a given $E_b$ and with  the potential 
$V_{\alpha\alpha}$ which is obtained from Eq. (\ref{Vgeneral}). This means 
that, if one chooses an approximate $\Psi$, it is still an exact solution of 
Eq. 
(\ref{Schreq}) with the corresponding approximate $V_{\alpha\alpha}$. The 
difference of this approximate $\alpha\alpha$ potential from the one, which 
is considered as the exact potential, characterizes the quality of the 
chosen $\Psi$. If one manages to find such a function $\Psi$ that the 
potential, 
which follows from Eq. (\ref{Vgeneral}), is almost the same as the exact one, 
one actually finds a reliable solution of Eq. (\ref{Schreq}) for a given energy 
$E_b$. This can be achieved by varying the parameters of a postulated function 
$\Psi$ in order to minimize the difference between the approximate and the 
exact $\alpha\alpha$ potentials.

\section{Two-body potentials}
\label{sec.potentials}
The main input information that is needed in the present work is given by 
the $\alpha\alpha$ and $\alpha n$ two-body potentials. As is explained in the 
next section, these potentials are needed for the $S$-wave and $P$-wave states, 
respectively. The first of them is 
taken from Ref. \cite{AliBodmer}. This is well known Ali-Bodmer 
$\alpha\alpha$-potential:
\begin{equation}
\label{alibodmerpot}
  V_{\alpha\alpha}(r)=
  V_Re^{-(\beta_Rr)^2}-V_Ae^{-(\beta_Ar)^2}+V_c(r)\ ,
\end{equation}
which involves the repulsive and attractive Gaussian terms as well as the 
repulsive electric (Coulomb-like) one,
\begin{equation}
\label{coulomb}
  V_c(r)=\frac{4e^2}{r}\mathop{\mathrm{erf}}
   \left(\frac{\sqrt{3}}{2R_\alpha}r\right ) \ .
\end{equation}
The parameters of this potential are: $V_R=1050$\,MeV, 
$\beta_R=0.8\,\mathrm{fm}^{-1}$, $V_A=150$\,MeV, 
$\beta_A=0.5\,\mathrm{fm}^{-1}$, $R_\alpha=1.44\,\mathrm{fm}$. The Coulomb term 
(\ref{coulomb}) takes into account a non-zero size of the $\alpha$ particle. 
The expression (\ref{coulomb}) is obtained under the assumption that its charge 
has a Gaussian distribution in space with the rms-radius $R_\alpha$.

The $\alpha n$ potential in the present work is the same that was suggested by 
Bang and Gignoux \cite{Bang}. It is of the Woods-Saxon type with a 
spin-orbit term:
\begin{equation}
\label{anpot}
  V_{\alpha n}(r)=\frac{W_1}{1+\exp\left[(r-R_1)/d_1\right]}+
  \frac{\vec{\ell}\cdot\vec{s}}{r}\frac{d}{dr}
  \frac{W_2}{1+\exp\left[(r-R_2)/d_2\right]}\ ,
\end{equation}
where $W_1=-43.0\,\mathrm{MeV}$, $R_1=2.0\,\mathrm{fm}$, 
$d_1=0.7\,\mathrm{fm}$, $W_2=40.0\,\mathrm{MeV\cdot \mathrm{fm}^2}$, 
$R_2=1.5\,\mathrm{fm}$, $d_2=0.35\,\mathrm{fm}$. For the quantum state with the 
total angular momentum $J=3/2$, orbital angular momentum $\ell=1$, and the 
neutron spin $s=1/2$, the action of the operator $\vec{\ell}\cdot\vec{s}$ is 
equivalent to multiplication by 1/2.

The potentials (\ref{alibodmerpot}, \ref{anpot}) were used by many authors in a 
number of calculations within the three-body models of ${}^9\mathrm{Be}$ 
and ${}^6\mathrm{Li}$ (see, for example, Refs. \cite{Efros, Filikhin, Bang} 
where the corresponding three-body Faddeev equations were solved). It turned 
out that these potentials reliably described the $\alpha\alpha$ and $\alpha n$ 
interactions at low energies as well as allowed one to well reproduce the 
binding energies and the other properties of the nuclei. The potentials  
(\ref{alibodmerpot}, \ref{anpot}) are therefore used in the present paper as 
well.

\section{Wave function of ${}^9\mathrm{Be}$}
\label{sec.wf}
As was explained in the Introduction, the Beryllium isotope ${}^9\mathrm{Be}$ is 
clusterized in two $\alpha$ particles and a neutron. This is a Borromean 
three-body system that has only one bound state with a rather small binding 
energy $|E_b|=1.5736\,\mathrm{MeV}$ and with the spin-parity $(3/2)^-$ 
\cite{Tilley}. Since the $\alpha$ particle spins are zero, the three-body wave 
function should be symmetric with respect to the $\alpha$-$\alpha$ permutation. 
This means that their relative orbital angular momentum is even, 
$\ell_{\alpha\alpha}=0,2,4,\dots$. The negative parity can therefore be obtained 
if the valence neutron moves with an odd orbital angular momentum, 
$\ell_n=1,3,5,\dots$, relative to the $\alpha\alpha$ pair.

As a reliable approximation, it is reasonable to assume that 
$\ell_{\alpha\alpha}=0$ and $\ell_n=1$.  All the higher partial waves are 
ignored since the $(\alpha\alpha n)$ binding is weak and Borromean while high 
values of $\ell_{\alpha\alpha}$ and  $\ell_n$ introduce additional repulsive 
centrifugal potentials into the three-body hamiltonianian. The minimal repulsion 
is in the state $\ell_{\alpha\alpha}=0$,  $\ell_n=1$. 

One may wonder why $\ell_n=1$ and not zero, which would generate even less 
repulsion. An intuitive explanation can be found by considering the naive shell 
model as follows. There are five neutrons in this nucleus. Four of them are 
sitting in the clusters. Each cluster has its own mean-field potential and its 
own level structure. Two neutrons in the first $\alpha$ particle occupy its 
lowest $S$-level and the other two neutrons occupy the lowest $S$-level in the 
second $\alpha$ particle. What remains for the fifth neutron is the $P$-level 
either in the first or in the second cluster. Therefore the valence neutron 
moves in the $P$-wave state with respect to both $\alpha$ particles. This is 
therefore the shell-model configuration with the minimal energy. And since 
$\ell_{\alpha\alpha}=0$, the orbital angular momentum of the neutron with 
respect of the centre of mass of the $\alpha\alpha$ subsystem is also one, i.e. 
$\ell_n=1$.

Based on the above reasoning, one comes to the following structure of the wave 
function of ${}^9\mathrm{Be}$:
\begin{equation}
\label{WF.structure}
   \Psi(\vec{r}_{\alpha\alpha},\vec{r}_n)=R(r_{\alpha\alpha},r_n)
   Y_{00}(\Omega_{\alpha\alpha})\mathcal{Y}_{1\frac12}^{\frac32J_z}
   (\Omega_{n})\ ,
\end{equation}
where the spin-angular function of the neutron, 
$\mathcal{Y}_{1\frac12}^{\frac32J_z}(\Omega_{n})$, defined as
\begin{equation}
\label{WF.Y}
   \mathcal{Y}_{\ell s}^{JJ_z}(\Omega)=
   \sum_{ms_z}C_{\ell mss_z}^{JJ_z}Y_{\ell m}(\Omega)\chi_s(s_z)\ ,
\end{equation}
couples its orbital angular momentum and the spin. Here $\chi_s$ is the spin 
function of the neutron.

Substituting wave function (\ref{WF.structure}) in Eqs. (\ref{Vgeneral}) 
and (\ref{D}), the following decomposition of the 
$\alpha\alpha$ potential is obtained:
\begin{equation}
\label{Vfinal}
   V_{\alpha\alpha}({r}_{\alpha\alpha})
   =E_b-
   \left\langle V_{\alpha n}(\rho_1)\phantom{\frac11}\!\!\!
   +V_{\alpha n}(\rho_2)\right\rangle
   -\left\langle\frac{\ell_n(\ell_n+1)}{r_n^2}\right\rangle
   + \left\langle E_{\mathrm{kin}}\right\rangle\ ,
\end{equation}
where the averaging, denoted by the brackets $\langle\rangle$, is done over all 
the configuration space variables except the radial variable $r_{\alpha\alpha}$.
These average values represent the contributions that are originated from the 
two $\alpha n$ potentials,
\begin{equation}
\label{Vaverage}
   \left\langle V_{\alpha n}(\rho_{1,2})\phantom{\frac11}\!\!\!\right\rangle
   =
   \frac{3}{4D(r_{\alpha\alpha})}\int_0^\infty\! dr_n\int_{0}^{\pi}\! 
   d\vartheta\,
   r_n^2(\sin\vartheta)^3 V_{\alpha n}(\rho_{1,2})
   \left|R(r_{\alpha\alpha},r_n)\right|^2\ ,
\end{equation}
from the centrifugal potential caused by $\ell_n=1$,
\begin{equation}
\label{Laverage}
   \left\langle\frac{\ell_n(\ell_n+1)}{r_n^2}\right\rangle=
   \frac{\hbar^2}{\mu D(r_{\alpha\alpha})}\int_0^\infty\! dr_n
   \left|R(r_{\alpha\alpha},r_n)\right|^2\ ,
\end{equation}
and from the radial derivatives (``kinetic energy'' terms) of the Laplacians
$\Delta_{\vec{r}_{\alpha\alpha}}$ and $\Delta_{\vec{r}_{n}}$,
\begin{eqnarray}
\nonumber
   \left\langle E_{\mathrm{kin}}\right\rangle
   &=&
   \frac{\hbar^2}{2\mu_{\alpha\alpha}r_{\alpha\alpha}^2
   D(r_{\alpha\alpha})}\int_0^\infty\! dr_n\,r_n^2R^*(r_{\alpha\alpha},r_n)
   \frac{\partial}{\partial r_{\alpha\alpha}}\left[r_{\alpha\alpha}^2
   \frac{\partial}{\partial r_{\alpha\alpha}}R(r_{\alpha\alpha},r_n)
   \right]\\[3mm]
\label{Eaverage}
   &+&
   \frac{\hbar^2}{2\mu 
   D(r_{\alpha\alpha})}\int_0^\infty\! dr_n\,R^*(r_{\alpha\alpha},r_n)
   \frac{\partial}{\partial r_{n}}\left[r_{n}^2
   \frac{\partial}{\partial r_{n}}R(r_{\alpha\alpha},r_n)
   \right]\ ,
\end{eqnarray}
where
\begin{equation}
\label{Daverage}
   D(r_{\alpha\alpha})=\int_0^\infty\! dr_n\,r_n^2
   \left|R(r_{\alpha\alpha},r_n)\right|^2\ .
\end{equation}
When obtaining Eq. (\ref{Vaverage}) it was used the fact that the left hand 
side of Eq. (\ref{Vfinal}) cannot depend on $J_z$. The simplest choice is  
$J_z=3/2$ which implies that the sum in Eq. (\ref{WF.Y}) only includes one 
term involving the spherical harmonics 
$Y_{11}(\vartheta,\varphi)=-(1/2)\sqrt{3/(2\pi)}\sin\vartheta\exp(i\varphi)$, 
where the $z$-axis is chosen along $\vec{r}_{\alpha\alpha}$ and the polar angle 
$\vartheta$ is between vectors $\vec{r}_{\alpha\alpha}$ and 
$\vec{r}_{n}$ (see Fig.~\ref{fig.Jacobi}).

It should be noted that in Eqs. (\ref{Vaverage}-\ref{Eaverage}) the 
normalization of the radial wave function $R(r_{\alpha\alpha},r_n)$ is 
arbitrary. This is because in each of these equations the square of $R$ is 
present in both the numerator and denominator. This fact significantly 
simplifies the procedure for fitting the potential 
$V_{\alpha\alpha}({r}_{\alpha\alpha})$, because one does not have to care about 
proper nomalization of $R(r_{\alpha\alpha},r_n)$ when varying its parameters.
The function $R(r_{\alpha\alpha},r_n)$ only needs to be normalized to unity at 
the final stage when all its parameters have been optimized and established.

The choice of a functional form of $R(r_{\alpha\alpha},r_n)$ is the most 
important and difficult part of the described procedure. It was done by trial  
and error approach. After many unsuccessful attempts, it was found such a form 
of this radial wave function that Eq. (\ref{Vfinal}) produced the potential 
curve whose shape was more or less similar to the Ali-Bodmer potential. Finally, 
the parameters of $R(r_{\alpha\alpha},r_n)$ were adjusted by minimizing the 
difference between the Ali-Bodmer potential and the one produced by Eq. 
(\ref{Vfinal}). This was done via the least-square method using the minimization 
code MINUIT \cite{MINUIT}.

The resulting radial wave function of ${}^9\mathrm{Be}$ looks as follows:
\begin{equation}
\label{WFresult}
   R(r_{\alpha\alpha},r_n)=\frac{N}{r_{\alpha\alpha}^2r_n^3}
   \arctan\left[(a_1r_{\alpha\alpha})^{7/2}\right]
   \arctan\left[(a_2r_{n})^{7/2}\right]
   \left[1-e^{-(a_3r_{\alpha\alpha})^2}\right]^2
   \frac{e^{-\varkappa\zeta}}{\zeta^{5/2}}\ ,
\end{equation}
where $\varkappa=\sqrt{2\mu|E_b|/\hbar^2}$ is the momentum corresponding to the 
experimental binding energy, 
\begin{equation}
\label{hyperr}
   \zeta=\sqrt{\frac{\mu_{\alpha\alpha}}{\mu}r_{\alpha\alpha}^2+
   r_n^2}
\end{equation}
is the hyperradius, 
$a_1=0.37441\,\mathrm{fm}^{-1}$, 
$a_2=0.070112\,\mathrm{fm}^{-1}$, 
$a_3=0.41953\,\mathrm{fm}^{-1}$, 
and the normalization constant $N=927446.734\,\mathrm{fm}^{9/2}$.

A specific choice of the functional form of $R(r_{\alpha\alpha},r_n)$ was based 
on some general properties of such a function as well as on the author's 
intuition. Unfortunately, there is no  universal recipe for constructing this 
function. 

The following general features of the wave function were taken 
into account. First of all, the radial wave function must exponentially tend 
to zero at large distances in all directions in the configuration space. It is 
known from the hyperspherical expansion theory that for any three-body system 
such an exponential diminishing is given by the factor 
$e^{-\varkappa\zeta}/\zeta^{5/2}$. This is why this factor is present in the 
expression (\ref{WFresult}). Due to the Pauli blocking, the $\alpha$ particles 
cannot penetrate each other. For the same reason the neutron cannot move inside 
them. Therefore the Pauli blocking requires vanishing of the wave 
function when $r_{\alpha\alpha}$ and $r_n$ tend to zero. And finally, the wave 
function should have maxima at some short distances along both radial variables, 
$r_{\alpha\alpha}$ and $r_n$. These maxima correspond to the most probable 
configuration of the system. 

A maximum at some intermediate distance along the variable $x$ as well as 
vanishing near $x=0$ can be implemented by using the function  
$\arctan[(ax)^n]/x^m$ with $n>m$. There are two such factors in the function 
(\ref{WFresult}), for $x=r_{\alpha\alpha}$ and for $x=r_{n}$. In principle, the 
powers $n$ and $m$ could be made additional adjustable parameters. However, it 
was decided not to increase the number of the free parameters, because a 
multi-variable function usually has many local minima and threfore with too 
many variable parameters the fitting of the potential could be more difficult. 
The optimal values of $n$ and $m$ for the variables $r_{\alpha\alpha}$ and 
$r_n$ were found manually by the trial and error method.

It turned out that the suppression of $R(r_{\alpha\alpha},r_n)$ at   
$r_{\alpha\alpha}\to0$ that was provided by the corresponding factor 
$\sim\arctan(.)$, was insufficient (the resulting $V_{\alpha\alpha}$ potential 
was too soft at the origin). In order to increase the suppression, an 
additional factor  $[1-e^{-(a_3r_{\alpha\alpha})^2}]^2$ was introduced.

\begin{figure}[ht!]
\centerline{\epsfig{file=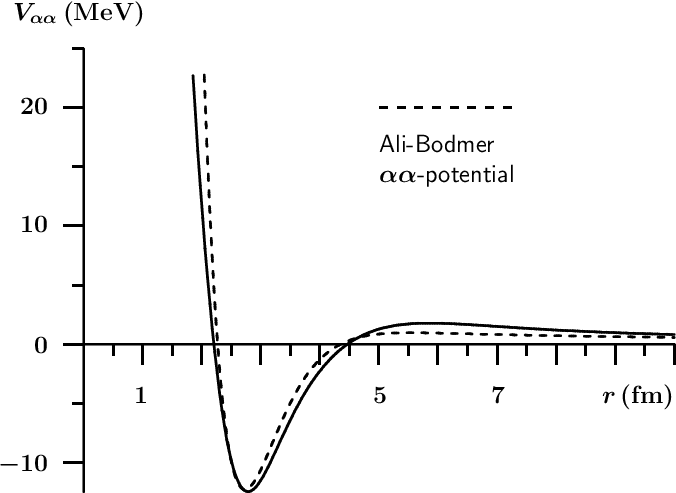}}
\caption{\sf
The exact Ali-Bodmer potential (\ref{alibodmerpot}) shown by the dashed curve, 
and the $\alpha\alpha$ potential corresponding to the wave function 
(\ref{WF.structure}) with the radial part (\ref{WFresult}) (solid curve).
}
\label{fig.fit}
\end{figure}

The wave function (\ref{WF.structure}) with the radial part (\ref{WFresult}) is 
an exact solution of Eq. (\ref{Vgeneral}) for the $\alpha\alpha$ potential 
shown in Fig. (\ref{fig.fit}) by the solid curve. As is seen, this 
reconstructed potential is very close to the Ali-Bodmer one. On one hand it is 
a bit different, but on the other hand it corresponds to the exact experimental 
binding energy of ${}^9\mathrm{Be}$.

\begin{figure}[ht!]
\centerline{\epsfig{file=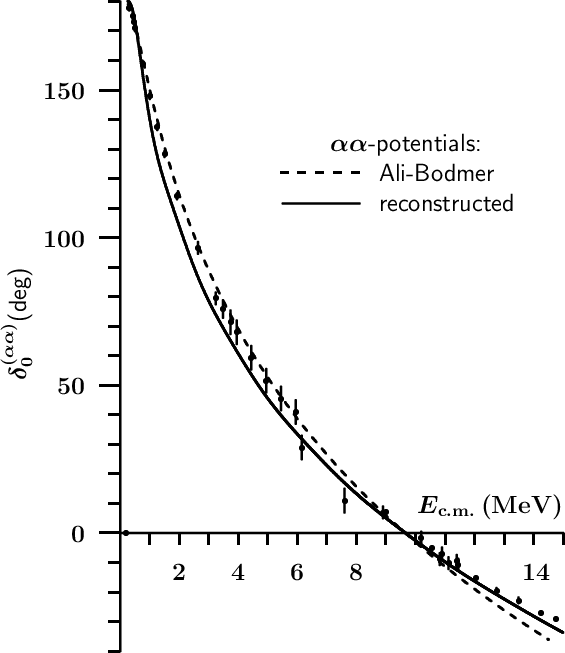}}
\caption{\sf
The $S$-wave phase shift for the Ali-Bodmer potential (\ref{alibodmerpot}) 
shown by the dashed curve, 
and the corresponding phase shift for the $\alpha\alpha$ potential 
reconstructed from the wave function 
(\ref{WF.structure}) with the radial part (\ref{WFresult}) (solid curve).
The experimental data are taken from Refs. \cite{Afzal, Chien}.
}
\label{fig.phase}
\end{figure}

In order to check how good the reconstructed potential is in describing the 
low-energy $\alpha\alpha$ scattering, the $S$-wave scattering phase shifts were 
calculated and compared with the corresponding values for the Ali-Bodmer 
potential. For obtaining the phase shifts the differential equations for the 
Jost functions were numerically integrated (see Eqs. (8.57, 8.58) of the book 
\cite{Rakityansky}). The results of these calculations are shown in Fig. 
\ref{fig.phase}, where the solid curve corresponds to the reconstructed 
potential while the dashed curve represents the Ali-Bodmer one. In Fig. 
\ref{fig.phase} these phase shifts are compared not only with each other but 
also with the corresponding experimental values. The experimental data are 
available (see Refs. \cite{Afzal, Chien}) as the so-called nuclear parts of the 
phase shifts, i.e. the pure Coulomb phase shifts are subtracted. In calculating 
the curves such a subtraction was done as well.

The purpose of this comparison is  to indirectly check the quality of 
the wave function (\ref{WF.structure}) with the radial part (\ref{WFresult}). 
Since the reconstructed potential is not very much different from the 
Ali-Bodmer one and since it generates almost the same phase shifts that are 
also close to experimental data, one can say that in a possible application of 
this wave function it can be considered as a reliable approximation of 
the exact solution of the three-body Schr\"odinger equation (\ref{Schreq}) with 
the two-body potentials given in Sec. \ref{sec.potentials}. 

An additional test of the wave function that is obtained in the present work 
can be done by calculating various space-distances in the 
bound $(\alpha\alpha n)$ state which this wave function describes. It should be 
emphasized that none of the root-mean-square (RMS) distances that are 
calculated and given in Table \ref{table.RMS}, were fitted. The fit (by varying 
the parameters $a_1$, $a_2$, and $a_3$) was only done for the $\alpha\alpha$ 
potential. The RMS distances are just those that the resulting wave function 
gives.

The calculation of the RMS distances is reduced to one- or three-dimensional 
integrals over $r_{\alpha\alpha}$, $r_n$, and $\vartheta$ (see Fig. 
\ref{fig.Jacobi}):
\begin{equation}
\label{raaAV}
   \langle r_{\alpha\alpha}^2\rangle=
   \left\langle\Psi\left|r_{\alpha\alpha}^2\right|\Psi\right\rangle\ ,
\end{equation}
\begin{equation}
\label{rcmnAV}
   \langle r_{\mathrm{cm}-n}^2\rangle=
   \left\langle\Psi\left|
   \left(\frac{2m_\alpha}{2m_\alpha+m_n}r_n\right)^2
   \right|\Psi\right\rangle\ ,
\end{equation}
\begin{equation}
\label{rcmaAV}
   \langle r_{\mathrm{cm}-\alpha}^2\rangle=
   \left\langle\Psi\left|
   \left(\frac12\vec{r}_{\alpha\alpha}-
   \frac{m_n}{2m_\alpha+m_n}\vec{r}_n\right)^2
   \right|\Psi\right\rangle\ .
\end{equation}
When finding the RMS charge and matter radii of ${}^9\mathrm{Be}$, the 
corresponding non-zero radii of the $\alpha$ particle, 
$R_{\mathrm{ch}(\alpha)}=1.67824\,\mathrm{fm}$ \cite{Nature} and 
$R_{\mathrm{mat}(\alpha)}=1.457\,\mathrm{fm}$ \cite{Tanihata}
should be taken into account:
\begin{equation}
\label{rchAV}
   \langle r_{\mathrm{ch}}^2\rangle=
   \langle r_{\mathrm{cm}-\alpha}^2\rangle+R_{\mathrm{ch}(\alpha)}^2\ ,
\end{equation}
\begin{equation}
\label{rmatAV}
   \langle r_{\mathrm{mat}}^2\rangle=
   \frac{2m_\alpha}{2m_\alpha+m_n}\left[
   \langle r_{\mathrm{cm}-\alpha}^2\rangle+R_{\mathrm{mat}(\alpha)}^2\right]+
   \frac{m_n}{2m_\alpha+m_n}\langle r_{\mathrm{cm}-n}^2\rangle\ .
\end{equation}

\begin{table}[ht]
\begin{center}
\begin{tabular}{|c|c|c|}
\hline
 distance   & calculated (fm) & measured (fm)\\
\hline
$\sqrt{\langle r_{\alpha\alpha}^2\rangle}$       & 3.46 &\\
\hline
$\sqrt{\langle r_{\mathrm{cm}-n}^2\rangle}$      & 4.65 &\\
\hline
$\sqrt{\langle r_{\mathrm{cm}-\alpha}^2\rangle}$ & 1.83 &\\
\hline
$\sqrt{\langle r_{\mathrm{ch}}^2\rangle}$        & 2.48 & 2.519 \cite{ADNDT1, 
                                                                 ADNDT2}\\
\hline
$\sqrt{\langle r_{\mathrm{mat}}^2\rangle}$  & 2.70 & $2.50\pm0.01$ \cite{PRL}\\
\hline
\end{tabular}
\end{center}
\caption{\sf
Various RMS distances in ${}^9\mathrm{Be}$: between the $\alpha$ particles 
($\sqrt{\langle r_{\alpha\alpha}^2\rangle}$), from c.m. to the valence neutron
($\sqrt{\langle r_{\mathrm{cm}-n}^2\rangle}$), from c.m. to the $\alpha$ 
particle 
($\sqrt{\langle r_{\mathrm{cm}-\alpha}^2\rangle}$), 
charge radius of the nucleus 
($\sqrt{\langle r_{\mathrm{ch}}^2\rangle}$),
matter radius of the nucleus 
($\sqrt{\langle r_{\mathrm{mat}}^2\rangle}$). The experimental data are taken 
from Refs. \cite{ADNDT1, ADNDT2, PRL}.
}
\label{table.RMS}
\end{table}

\begin{figure}[ht!]
\centerline{\epsfig{file=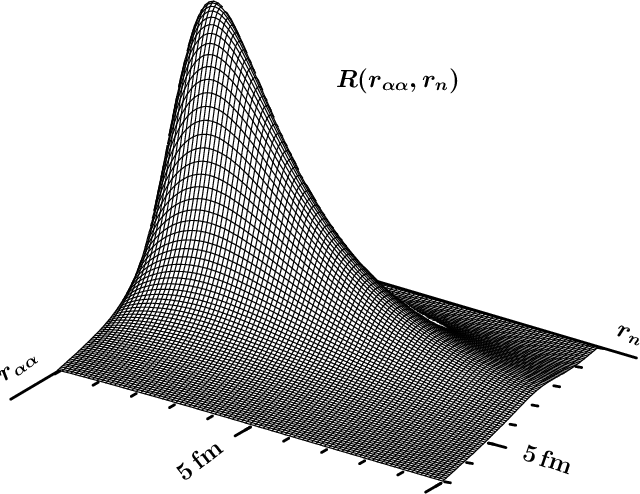}}
\caption{\sf
Radial part (\ref{WFresult}) of the wave function of ${}^9\mathrm{Be}$.
}
\label{fig.3D}
\end{figure}

As is seen, the charge radius, $\sqrt{\langle r_{\mathrm{ch}}^2\rangle}$, turned 
out to be very close to the corresponding experimental value. The 
matter radius, $\sqrt{\langle r_{\mathrm{mat}}^2\rangle}$, is a bit too large 
as 
compared to the measurement. The difference, however, is not drastic. A very 
close result, namely, the value of 2.68\,fm for the matter radius of 
${}^9\mathrm{Be}$ was obtained in Ref. \cite{Theeten} within the three-body 
model with microscopic nonlocal interactions based on Volkov V2 potential. Also, 
it should be noted that in contrast to a rather simple and accurate procedure 
for experimental determination of the charge radius, the 
measurements of the 
matter radii
of the nuclei are always more difficult and 
are not without some ambiguities \cite{Tanihata}.

A quasi-3D image of the radial part (\ref{WFresult}) of the 
wave function of ${}^9\mathrm{Be}$ is given in Fig. \ref{fig.3D}. This function 
has a maximum at $r_{\alpha\alpha}=2.4051\,\mathrm{fm}$ and 
$r_{n}=1.4654\,\mathrm{fm}$. It is also clearly seen a ``ridge'' along $r_n$ 
with $r_{\alpha\alpha}\sim2\,\mathrm{fm}$. This extended ``ridge'' is the cause 
of rather large values of $\sqrt{\langle r_{\mathrm{cm}-n}^2\rangle}$ and 
$\sqrt{\langle r_{\mathrm{mat}}^2\rangle}$.

\section{Conclusion}
\label{concl}
The wave function (\ref{WF.structure}) with the radial part (\ref{WFresult}) is 
an exact bound-state solution of the three-body integro-differential equation 
(\ref{Vgeneral}) which follows from the equivalent  Schr\"odinger 
equation (\ref{Schreq}). By construction, this solution corresponds to the 
experimentally known binding energy, as well as to the $\alpha n$ potential 
(\ref{anpot}), and to the $\alpha\alpha$ potential that is almost the same as 
the Ali-Bodmer one. It can therefore be considered as a reliable approximation 
of the wave function of the nucleus ${}^9\mathrm{Be}$ in the three-body model 
($\alpha\alpha n$).

This wave function is obtained in the form of a compact analytical 
expression that is easy to use. Among possible applications of this function 
could be, for example, the constructing of various folding potentials as well 
as estimating the cross sections of various nuclear reactions that involve 
${}^9\mathrm{Be}$.

The method used in the present paper, is universal and can be applied for 
constructing the ground-state wave functions of any other nuclear or some 
atomic bound three-body systems. Of course it cannot compete with the approaches 
based on Faddeev equations, hyperspherical expansions, etc., where whole 
spectrum of the excited states can be described. However, the proposed method 
has its own application niche where it may be preferable. In particular, this 
method is suitable for the systems that have no excited states, such as, for 
example, the nucleui ${}^6\mathrm{He}$ and ${}^6\mathrm{Li}$ if they are 
considered in the three-body model $(\alpha NN)$. The proposed method can give 
an analytical expression for the wave function which is easy to use when one 
needs a reliable estimate of something and does not want to delve into 
complicated numerical calculations of the rigorous three-body theory. In 
principle, the excited state wave functions can be constructed within this 
method as well. This, however, will require additional effort in making the 
excited states orthogonal to the ground state and to each other.

The method can be generalized for constructing the wave functions that are 
linear combinations of several mutually orthogonal components. For example, the 
ground $0^+$-state of ${}^6\mathrm{He}$ in the $(\alpha nn)$-model is mainly 
composed of the two components: the state $|\Psi_0\rangle$ with zero $nn$-spin 
as well as zero orbital angular momenta associated with both $nn$ and 
$\alpha(nn)$ Jacobi coordinates (82.87\%), plus the state $|\Psi_1\rangle$ with 
the  $nn$-spin one and the $P$-waves along both Jacobi coordinates (13.96\%) 
\cite{Danilin}, i.e. $|\Psi\rangle=C_0|\Psi_0\rangle + C_1|\Psi_1\rangle$. In 
such a case, when going from Eq. (\ref{Schreq}) (which should be written for 
the $nn$-potential) to Eq. (\ref{Vgeneral}), one should multiply Eq. 
(\ref{Schreq}) from the left either by $\Psi_0^\dagger$ or by $\Psi_1^\dagger$ 
and integrate over all the configuration-space variables except the 
$nn$-distance. 
As a result one obtains a system of two coupled integro-differential equations 
where on the left hand sides are the singlet (${}^1S_0$) and triplet (${}^3P_1$)
$nn$-potentials which can be fitted by adjusting the parameters of some 
postulated $\Psi_0$ and $\Psi_1$.

Another possible generalization of the proposed method may consist in 
considering four-, five-, and larger numbers of bodies in the bound system. In 
doing this, one term of the Schr\"odinger equation that involves a chosen 
two-body potential, can be moved to the left-hand side  as is done in Eq. 
(\ref{Schreq}). Then whole equation should be multiplied from the left by 
$\Psi^\dagger$ and integrated over all the variables except the one which the 
chosen potential depends on. After that the wave function should be postulated 
and its parameters be adjusted in order to fit the potential. The main 
difficulty here is that with increasing number of particles it is not a simple 
task  to find (to guess) an appropriate parametrization of the wave function 
because it depends on many variables. Moreover, even if one manages to find 
such a 
parametrization, the right-hand side of Eq. (\ref{Vgeneral}) will involve 
multi-dimensional integrals which may cause some numerical difficulties when the 
system under consideration consists of too many particles.



\begin{thebibliography}{99}
\bibitem{Tang}
   Y.C. Tang, F.C. Khanna, R.C. Herndon, K. Wildermuth,
   ``$\mathrm{Be}^9$ in the cluster model'', 
   Nucl. Phys., {\bf 35}, 421-433 (1962)
\bibitem{Efros}   
   V.D. Efros, H. Oberhummer, A. Pushkin, I.J. Thompson,
   ``Low-energy photodisintegration of ${}^9\mathrm{Be}$ and 
   $\alpha+\alpha+n \leftrightarrow {}^9\mathrm{Be}+\gamma$ reactions at 
   astrophysical conditions'',
   The European Physical Journal A - Hadrons and Nuclei, {\bf 1(4)}, 447-453 
   (1998)\\ doi:10.1007/s100500050079
\bibitem{Filikhin}
   I. Filikhin, V.M. Suslov, B. Vlahovic,
   ``$\mathrm{Be}^9$ Low-Lying Spectrum Within a Three-Cluster Model,
   Few-Body Syst., {\bf 50}, 255-257 (2011)\\ 
   https://doi.org/10.1007/s00601-010-0135-3
\bibitem{Casal}   
   J. Casal, M. Rodrıiguez-Gallardo, J.M. Arias, and I.J. Thompson, 
   ``Astrophysical reaction rate for ${}^9\mathrm{Be}$ formation within a 
   three-body approach'',
   Phys.Rev. C, {\bf 90}, 044304 (2014)\\
   DOI: 10.1103/PhysRevC.90.044304
\bibitem{Filandri}
   Elena Filandri, Paolo Andreatta, Carlo A. Manzata, Chen Ji, W. 
   Leidemann,  and G. Orlandini,
   ``Beryllium-9 in Cluster Effective Field Theory'',
   arXiv:2002.00780v1 [nucl-th] (2020)
\bibitem{Tilley}
   D.R. Tilley, J.H. Kelley, J.L. Godwin, D.J. Millener, J.E. Purcell, C.G. 
   Sheu, and H.R. Weller, 
   ``Energy levels of light nuclei A = 8, 9, 10'',
   Nucl. Phys., A {\bf 745}, 155-362 (2004)
\bibitem{cauldrons}
   Claus E. Rolfs and William S. Rodney,
   ``Cauldrons in the Cosmos'', The University of Chicago Press (1988)
\bibitem{Aprahamian}
   A. Aprahamian, K. Langanke, and M. Wiescher, 
   ``Nuclear structure aspects in nuclear astrophysics'', 
   Prog. Part. Nucl. Phys., {\bf 54}, 535-613 (2005)
\bibitem{Khan}
   Dr. Md A Khan,
   ``Nuclear Astrophysics - A Course of Lectures'',
   CRC Press, Taylor \& Francis Group, 
   International Standard Book Number-13: 978-1-138-58816-5, (2018)
\bibitem{Mohr}
   Peter Mohr,
   ``Direct capture cross section of 
    ${}^9\mathrm{Be}(n,\gamma){}^{10}\mathrm{Be}$'',
    Phys. Rev. C, {\bf 99}, 055807 (2019)
\bibitem{Carbone}
   D. Carbone, M. Bondi,  A. Bonaccorso, C. Agodi,  F. Cappuzzello, M. 
   Cavallaro,  R. J. Charity, A. Cunsolo, M. De Napoli, and A. Foti,
   ``First application of the $n-{}^9\mathrm{Be}$ optical potential to the 
    study of the ${}^{10}\mathrm{Be}$ continuum via the 
    $({}^{18}\mathrm{O},{}^{17}\mathrm{O})$ neutron-transfer reaction'',
    Phys. Rev. C, {\bf 90}, 064621 (2014)
\bibitem{SuJun}    
    Su Jun, Li Zhi-Hong, Guo Bing, Liu Wei-Ping, Bai Xi-Xiang, Zeng Sheng, 
    Lian Gang, Yan Sheng-Quan, Wang Bao-Xiang, and Wang You-Bao,
    ``Astrophysical Reaction Rates of the 
    ${}^8\mathrm{Li}(p,\gamma){}^{9}\mathrm{Be}_{\mathrm g.s.}$ Direct Capture 
    Reaction'',
    Chinese Physics Letters, {\bf 23 (1)}, 55-57 (2006)
\bibitem{Dub}    
    S.B. Dubovichenko, N.A. Burkova, A.V. Dzhazairov-Kakhramanov,
    ``The role of resonances in the capture of 
    ${}^8\mathrm{Li}(p,\gamma){}^{9}\mathrm{Be}$ on 
    the reaction rate of the relevant astrophysical synthesis of 
    ${}^9\mathrm{Be}$'',
    Nucl. Phys. A, {\bf 1000},121842 (2020) 
    https://doi.org/10.1016/j.nuclphysa.2020.121842
\bibitem{Itagaki}
    N. Itagaki, and K. Hagino,
    ``Low-energy photodisintegration of ${}^9\mathrm{Be}$ with the molecular 
    orbit model'',
    Phys. Rev. C, {\bf 66}, 057301 (2002)
\bibitem{Descouvemont1}
    P. Descouvemont, T. Druet, L.F. Canto, M.S. Hussein,
    ``Low-energy ${}^9\mathrm{Be}+{}^{208}\mathrm{Pb}$ scattering, breakup, and 
    fusion within a four-body model'',
    Phys. Rev. C,  {\bf 91}, 024606 (2015)
    DOI: 10.1103/PhysRevC.91.024606
\bibitem{Descouvemont}
    P. Descouvemont, and N. Itagaki,
    ``${}^9\mathrm{Be}$ scattering with microscopic wave functions and the 
    continuum-discretized coupled-channel method'',
    Phys. Rev. C, {\bf97}, 014612 (2018)
    DOI: 10.1103/PhysRevC.97.014612
\bibitem{Belyaev}
    V.B. Belyaev, S.A. Rakityansky, I.M. Gopane,
    ``Recovering the Two-Body Potential from a Given Three-Body Wave 
    Function'', 
    Few-Body Syst., {\bf 64}, 4 (2023) 
    https://doi.org/10.1007/s00601-022-01785-7
\bibitem{AliBodmer}
    S. Ali and A.R. Bodmer,
    ``PHENOMENOLOGICAL alpha-alpha POTENTIALS'',
    Nucl.Phys.,{\bf 80}, 99-112 (1966)
\bibitem{Bang}
    J. Bang, C. Gignoux,
    ``A REALISTIC THREE-BODY MODEL OF ${}^6\mathrm{Li}$ WITH LOCAL 
    INTERACTIONS'',
    Nucl. Phys. A, {\bf 313}, 119-140 (1979)
\bibitem{Thompson}
    I.J. Thompson, B.V. Danilin, V.D. Efros, J.S. Vaagen,
    J.M. Bang, M.V. Zhukov,
    ``Pauli blocking in three-body models of halo nuclei'',
    Phys.Rev. C, {\bf 61} 024318 (2000)
\bibitem{MINUIT}
        F. James and M. Roos,
        ``Minuit - a system for function minimization and analysis of the
         parameter errors and correlations'',
        Comp. Phys. Comm. 10 (1975) 343-367; http://hep.fi.infn.it/minuit.pdf
\bibitem{Rakityansky}
    S. A. Rakitiansky, ``Jost Functions in Quantum Mechanics: A Unified 
    Approach to Scattering, Bound, and Resonant State Problems'', 
    Springer (2022)
\bibitem{Afzal}
    S.A. Afzal, A.A.Z. Ahmad, S. Ali,
    ``Systematic Survey of the $\alpha$-$\alpha$ Interaction'',
    Rev. Mod. Phys., {\bf 41}(1), 247-273 (1969)
\bibitem{Chien}
    W.S. Chien and Ronald E. Brown,
    ``Study of the $\alpha$+$\alpha$ system below 15\,MeV (c.m.)'',
    Phys. Rev. {\bf C10}(5), 1767-1784 (1974)    
\bibitem{Nature}
    Wilfried Nortershauser,
    ``Helium nucleus measured with record precision'',
      Nature, {\bf 589}, 518-519 (2021)
\bibitem{Tanihata}
      I. Tanihata, H. Savajols, R. Kanungo,
      ``Recent experimental progress in nuclear halo structure studies'',
      Progress in Particle and Nuclear Physics, 
      {\bf 68},215-313 (2013)     
\bibitem{ADNDT1}
    I. Angeli, K.P. Marinova,
    ``Table of experimental nuclear ground state charge radii: An update'',
    At. Data and Nucl. Data Tables, {\bf 99}(1), 69-95 (2013)\\
    https://doi.org/10.1016/j.adt.2011.12.006
\bibitem{ADNDT2}
    Tao Li, Yani Luo, Ning Wang, 
    ``Compilation of recent nuclear ground state charge radius measurements and 
    tests for models'',
    At. Data and Nucl. Data Tables, {\bf 140}, 101440 (2021)\\
    https://doi.org/10.1016/j.adt.2021.101440
\bibitem{PRL}
    I. Tanihata, H. Hamagaki, O. Hashimoto, Y. Shida, and N. Yoshikawa,
    ``Measurements of Interaction Cross Sections and Nuclear Radii in the Light 
    $p$-Shell Region'',
    Phys.Rev.Lett., {\bf 55}(24), 2676-2679 (1985)
\bibitem{Theeten} 
    M. Theeten, H. Matsumura, M. Orabi, D. Baye, P. Descouvemont, Y. Fujiwara, 
    and Y. Suzuki, 
    ``Three-body model of light nuclei with microscopic nonlocal 
    interactions'', 
    Phys. Rev., {\bf C76}, 054003 (2007)
\bibitem{Danilin}
    B.V. Danilin and M.V. Zhukov, S.N. Ershov, F.A. Gareev, and R.S. 
    Kurmanov, J.S. Vaagen, J.M. Bang,
    ``Dynamical multicluster  model for electroweak and charge-exchange
    reactions'',
    Phys. Rev., {\bf C43}(6), pp. 2835-2843 (1991)
\end{thebibliography}
\end{document}